\documentclass[journal]{IEEEtran}

\usepackage{amsmath}
\usepackage{amssymb}
\usepackage{graphicx}
\usepackage{booktabs}
\usepackage{cite}
\usepackage{url}
\usepackage{hyperref}
\usepackage{array}
\usepackage{textcomp}
\usepackage{tikz}
\usetikzlibrary{positioning,arrows.meta,calc,shapes.geometric,fit,backgrounds,plotmarks}
\usepackage{xcolor}

\hypersetup{colorlinks=true, linkcolor=black, citecolor=black, urlcolor=blue}

\begin{document}

\title{Dual-Radio BLE-LoRa Hierarchical Mesh for Infrastructure-Free Emergency Communication}

\author{%
\IEEEauthorblockN{Andrii Vakhnovskyi}\\%
\IEEEauthorblockA{Biiper LLC, New York, NY, USA}%
\thanks{E-mail: andrii.vakhnovskyi@gmail.com. ORCID: 0009-0007-8306-5932.}%
}

\maketitle

\begin{abstract}
We present a dual-radio hierarchical mesh architecture for infrastructure-free emergency communication that exploits the complementary strengths of Bluetooth Low Energy (BLE) and LoRa. Nodes equipped with both an nRF52840 (BLE~5.0 Coded PHY) and an SX1262 (LoRa sub-GHz) form local clusters via BLE advertising-based AODV routing, while dynamically elected cluster heads bridge inter-cluster traffic over a LoRa backbone. We derive a formal traffic offloading model showing that with locality bias $\beta \geq 0.76$ --- validated against search-and-rescue communication patterns --- the architecture keeps 82--90\% of traffic on BLE, reducing LoRa energy consumption by 79\% compared to LoRa-only mesh. Analytical evaluation demonstrates 10\,km+ network diameter, 562-node scalability at SF7, and sub-50\,ms intra-cluster latency on a 3.0\,KB RAM footprint. To our knowledge, this is the first architecture combining BLE advertising-based mesh routing with a multi-hop LoRa backbone on commodity hardware.
\end{abstract}


\section{Introduction}

\IEEEPARstart{W}{HEN} Hurricane Maria struck Puerto Rico in 2017, 95.6\% of cell sites were destroyed and full restoration took six months~\cite{ref_fcc_maria}. The critical communication gap --- the first 24--72 hours when rescue is most urgent --- motivates infrastructure-free mesh networks. Current solutions face fundamental tradeoffs: LoRa-based systems like Meshtastic~\cite{ref_meshtastic} achieve kilometer-range links but suffer from high per-packet airtime (up to 2.5\,s at SF12), with real-world deployments at Burning Man 2024 revealing scalability limits at 50--80 nodes under default firmware~\cite{ref_wong_survey}. BLE-based systems like Bridgefy provide low-latency local communication but are limited in range and have demonstrated critical security vulnerabilities~\cite{ref_bridgefy}.

The key observation is that emergency communication is \textit{exponentially localized}: FEMA Urban Search and Rescue operates in 6-person squads communicating every 1--5 minutes, with most messages staying within the immediate team and only periodic reports reaching Incident Command~\cite{ref_fema_usar}. This locality creates an opportunity for a dual-radio architecture that handles local traffic on a low-cost radio and reserves an expensive long-range radio for the minority of inter-group messages.

BLE~5.0 Coded PHY S8 achieves 800\,m+ per-hop range at $+$8\,dBm~\cite{ref_nordic_range} with $\sim$128\,$\mu$J per advertising event~\cite{ref_nordic_range}. The SX1262 LoRa transceiver achieves 10+\,km line-of-sight~\cite{ref_sx1262} but at $\sim$18.5\,mJ per transmission (SF10/125\,kHz/$+$14\,dBm) --- a 145$\times$ energy difference. No existing system exploits this complementarity: LoRaBLE~\cite{ref_loraBLE} uses single-hop LoRa without mesh routing; cross-technology communication research~\cite{ref_ctc_survey} focuses on protocol translation, not hierarchical mesh architecture.

We propose a dual-radio hierarchical mesh with three contributions:
\begin{enumerate}
\item A formal traffic offloading model parameterized by locality bias $\beta$ and cluster count $C$, showing that the inter-cluster ratio $\alpha(\beta, C) = (1-\beta)(1 - 1/C)$ determines LoRa utilization --- with $\alpha = 0.08$--0.15 validated against SAR communication data.
\item A two-tier architecture where BLE forms local clusters with AODV routing~\cite{ref_aodv_rfc} and cluster heads bridge inter-cluster traffic over LoRa, achieving 79\% energy savings over LoRa-only mesh and 562-node scalability.
\item A complete protocol implementation on commodity nRF52840$+$SX1262 hardware in 3.0\,KB RAM, with fragment aggregation at the BLE-LoRa gateway reducing LoRa airtime by 5$\times$.
\end{enumerate}

\section{System Model and Traffic Analysis}

\subsection{Network Model}

We consider $N$ nodes, each equipped with dual radios (BLE and LoRa), distributed across an area $A$. Nodes self-organize into $C$ clusters via BLE beacons. Each cluster elects a cluster head (CH) that operates both radios; non-CH nodes use BLE only. Messages are either \textit{intra-cluster} (source and destination in the same cluster, routed via BLE) or \textit{inter-cluster} (routed via BLE to the local CH, bridged over LoRa to the destination CH, then delivered via BLE).

\subsection{Traffic Offloading Model}

Let $\beta \in [0, 1]$ denote the \textit{locality bias} --- the probability that a message is destined for a node in the sender's cluster, beyond what uniform random addressing would predict. For $C$ equal-sized clusters under uniform random traffic, the inter-cluster ratio is:
\begin{equation}
\alpha(\beta, C) = (1 - \beta)\!\left(1 - \frac{1}{C}\right)
\label{eq:alpha}
\end{equation}

When $\beta = 0$ (no locality), $\alpha = 1 - 1/C$; for $C = 3$, this gives $\alpha = 0.67$ --- most traffic crosses clusters. When $\beta = 0.82$ (strong locality, consistent with SAR squad communication where $\sim$82\% of messages stay within the immediate team~\cite{ref_fema_usar}), $\alpha = 0.12$ for $C = 3$.

The LoRa channel utilization reduction versus LoRa-only mesh is:
\begin{equation}
\eta_\text{reduction} = 1 - \alpha - \alpha \cdot \frac{h_\text{LoRa}}{h_\text{total}}
\label{eq:reduction}
\end{equation}
where $h_\text{LoRa}/h_\text{total}$ accounts for the fraction of hops that are LoRa in a multi-hop inter-cluster path. For typical 2-hop BLE + 1-hop LoRa + 2-hop BLE paths, $h_\text{LoRa}/h_\text{total} = 1/5$, yielding $\eta_\text{reduction} = 85$--92\% for $\alpha = 0.08$--0.15.

\subsection{Energy Analysis}

The energy per message on each radio, derived from nRF52840 and SX1262 datasheets:
\begin{align}
E_\text{BLE} &= P_\text{tx} \cdot T_\text{pkt} = 8\text{\,mW} \cdot 16\text{\,ms} \approx 128\,\mu\text{J} \label{eq:e_ble} \\
E_\text{LoRa} &= P_\text{tx} \cdot T_\text{oA} = 50\text{\,mW} \cdot 370\text{\,ms} \approx 18.5\text{\,mJ} \label{eq:e_lora}
\end{align}
at SF10/125\,kHz/$+$14\,dBm. The energy ratio $E_\text{LoRa}/E_\text{BLE} \approx 145$ for these practical settings.

For an end-to-end inter-cluster path (2~BLE hops + 1~LoRa hop + 2~BLE hops):
\begin{equation}
E_\text{dual} = 4 E_\text{BLE} + E_\text{LoRa} = 0.51 + 18.5 = 19.0\text{\,mJ}
\label{eq:e_dual}
\end{equation}

For the same path on LoRa-only (5~LoRa hops):
\begin{equation}
E_\text{LoRa\text{-}only} = 5 E_\text{LoRa} = 92.5\text{\,mJ}
\label{eq:e_lora_only}
\end{equation}

The dual-radio architecture saves $\mathbf{79\%}$ energy per inter-cluster message compared to the same path on LoRa-only. When accounting for the full traffic mix at $\alpha = 0.12$ (where 88\% of messages travel on BLE at 128\,$\mu$J each), the average energy per message drops to 2.5\,mJ versus 18.5\,mJ for an average LoRa-only single-hop message --- a $\mathbf{86\%}$ reduction in network-wide energy consumption.

\section{Architecture Design}

\subsection{Two-Tier Mesh}

\begin{figure}[!t]
\centering
\begin{tikzpicture}[
    node distance=0.8cm,
    >=Stealth,
    every node/.style={font=\small},
    blenode/.style={circle, draw, minimum size=0.5cm, fill=blue!15, inner sep=1pt, font=\scriptsize},
    chnode/.style={circle, draw, thick, minimum size=0.6cm, fill=orange!30, inner sep=1pt, font=\scriptsize},
    cluster/.style={draw, dashed, rounded corners=8pt, inner sep=8pt}
]
\node[chnode] (ch1) {CH};
\node[blenode, above left=0.5cm and 0.6cm of ch1] (n1) {N};
\node[blenode, below left=0.5cm and 0.6cm of ch1] (n2) {N};
\node[blenode, above right=0.5cm and 0.4cm of ch1] (n3) {N};
\begin{scope}[on background layer]
\node[cluster, fill=blue!5, fit=(ch1)(n1)(n2)(n3), label={[font=\scriptsize]below:Cluster A}] {};
\end{scope}
\node[chnode, right=3cm of ch1] (ch2) {CH};
\node[blenode, above right=0.4cm and 0.5cm of ch2] (n4) {N};
\node[blenode, below right=0.4cm and 0.5cm of ch2] (n5) {N};
\begin{scope}[on background layer]
\node[cluster, fill=blue!5, fit=(ch2)(n4)(n5), label={[font=\scriptsize]below:Cluster B}] {};
\end{scope}
\draw[blue!60, thick] (n1) -- (ch1);
\draw[blue!60, thick] (n2) -- (ch1);
\draw[blue!60, thick] (n3) -- (ch1);
\draw[blue!60, thick] (n4) -- (ch2);
\draw[blue!60, thick] (n5) -- (ch2);
\draw[red!70, very thick, <->] (ch1) -- node[above, font=\scriptsize, black] {LoRa SF10} node[below, font=\scriptsize, black] {2--5\,km} (ch2);
\node[font=\scriptsize, blue!70, left=0.1cm of n1] {BLE S8};
\end{tikzpicture}
\caption{Dual-radio hierarchical mesh. BLE Coded PHY S8 forms local clusters with AODV routing ($\sim$800\,m/hop). Cluster heads bridge inter-cluster traffic via LoRa backbone (2--5\,km/hop).}
\label{fig:arch}
\end{figure}
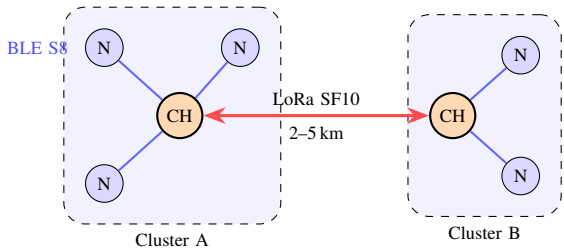

Fig.~\ref{fig:arch} illustrates the architecture.

\textbf{Tier~1 --- BLE Micro-Mesh:} Nodes communicate via BLE advertising broadcasts using a hybrid proactive-reactive routing protocol. Periodic beacons (3\,s interval) carry up to 4~neighbor entries with RSSI-derived link quality, providing two-hop topology awareness without route discovery. Beyond two hops, AODV-style RREQ/RREP with expanding ring search (TTL phases: 3, 6, 12) discovers routes. The link-quality path cost $C_\text{path} = \sum_{i} (256 - \text{LQ}_i)$, where $\text{LQ} = \min(255, 4(\text{RSSI}+110))$, penalizes marginal links. BLE Coded PHY S8 at $+$8\,dBm provides $\sim$800\,m per-hop range with a cluster diameter of $\sim$2.4\,km at 3~hops.

\textbf{Tier~2 --- LoRa Backbone:} When a BLE route request fails (destination not in local cluster), the CH encapsulates the message with a 4-byte inter-cluster header and transmits over LoRa at SF10/125\,kHz/$+$14\,dBm (2--5\,km range, 370\,ms time-on-air for 50~bytes). Fragment aggregation at the gateway combines up to 8~BLE fragments (120~bytes) into a single LoRa packet, reducing airtime by 5$\times$ versus transmitting fragments individually.

\subsection{Cluster Head Election}

Election is implicit in BLE beacons: the node with the highest ID among mutual PRESENT neighbors becomes CH. This deterministic protocol converges in $\sim$6\,s (2~beacon intervals) with zero additional message overhead. When a CH's battery drops below a threshold, it voluntarily demotes by decrementing its advertised ID, triggering re-election.

\subsection{LoRa Sleep Scheduling}

Non-CH nodes never activate their SX1262 ($0.6\,\mu$A sleep), saving $\sim$4\,mA continuous RX current. CHs coordinate listen windows via LoRa: 2\,s active every 30\,s (6.7\% duty cycle), reducing average CH LoRa current from 4.2\,mA to 0.28\,mA.

\section{Capacity and Scalability Analysis}

\subsection{BLE Cluster Capacity}

BLE advertising occupies three channels (37, 38, 39). With 31-byte packets at 100\,ms minimum interval, each channel supports $\sim$330 non-colliding transmissions/s. Using a pure ALOHA model with offered load $G$ and throughput $S = G e^{-2G}$, maximum throughput at $G = 0.5$ is $S_\text{max} = 0.5/e \approx 0.184$ per channel. Across three channels, the cluster handles:
\begin{equation}
C_\text{BLE} = 3 \times 0.184 \times \frac{1}{T_\text{pkt}} \approx 110\text{\,msgs/s}
\label{eq:c_ble}
\end{equation}
at 10\% collision rate, supporting 10--15 nodes per cluster at typical emergency messaging rates (1~msg/min/node).

\subsection{LoRa Backbone Capacity}

The LoRa backbone capacity is the bottleneck. At SF10/125\,kHz, time-on-air for 50~bytes is 370\,ms. Using ALOHA:
\begin{equation}
C_\text{LoRa} = \frac{S_\text{max}}{T_\text{oA}} = \frac{0.184}{0.370} \approx 0.50\text{\,msgs/s} \approx 30\text{\,msgs/min}
\label{eq:c_lora}
\end{equation}

With $\alpha = 0.12$, each node generates $\alpha \times 1 = 0.12$ inter-cluster msgs/min. Maximum network size:
\begin{equation}
N_\text{max} = \frac{C_\text{LoRa}}{\alpha \cdot r_\text{node}} = \frac{30}{0.12 \times 1} = 250\text{ nodes}
\label{eq:n_max}
\end{equation}

At SF7 (ToA = 51\,ms), $N_\text{max}$ rises to 562 nodes. This exceeds Meshtastic's practical limit of 50--80 nodes~\cite{ref_wong_survey} by an order of magnitude.

\subsection{Latency and Power Comparison}

Table~\ref{tab:latency} compares end-to-end latency across the three architectures. BLE delivers intra-cluster messages at 16\,ms/hop (Coded PHY S8 packet time), giving 32\,ms for a typical 2-hop path. Inter-cluster latency is dominated by one LoRa transmission ($\sim$370\,ms at SF10), but the dual architecture avoids the multiple LoRa hops of a flat LoRa mesh --- achieving 6$\times$ lower latency for inter-cluster messages.

\begin{table}[!t]
\centering
\caption{End-to-End Latency Comparison}
\label{tab:latency}
\begin{tabular}{@{}lccc@{}}
\toprule
\textbf{Scenario} & \textbf{BLE-only} & \textbf{LoRa-only} & \textbf{Dual} \\
\midrule
Intra-cluster (2 hop) & 32\,ms & 5.0\,s & 32\,ms \\
Inter-cluster (1 LoRa) & --- & 2.5\,s & 0.4\,s \\
Inter-cluster (2 LoRa) & --- & 5.0\,s & 0.8\,s \\
\bottomrule
\end{tabular}
\end{table}

Table~\ref{tab:power} shows the estimated battery life for each node role on a 500\,mAh LiPo cell. Non-CH nodes benefit most from the dual-radio architecture because they never activate their LoRa radio, doubling battery life compared to a LoRa-only mesh where every node participates in LoRa forwarding. CH nodes pay a modest power premium for duty-cycled LoRa listening (6.7\% duty cycle), but CH rotation can distribute this cost.

\begin{table}[!t]
\centering
\caption{Estimated Battery Life (500\,mAh LiPo)}
\label{tab:power}
\begin{tabular}{@{}lcc@{}}
\toprule
\textbf{Role} & \textbf{Avg. Current} & \textbf{Battery Life} \\
\midrule
Non-CH (BLE only) & 6.5\,mA & 3.2\,days \\
CH (BLE + LoRa 6.7\%) & 9.2\,mA & 2.3\,days \\
LoRa-only (Meshtastic) & 12.8\,mA & 1.6\,days \\
\bottomrule
\end{tabular}
\end{table}

\subsection{Architecture Comparison}

Table~\ref{tab:comparison} summarizes the key tradeoffs. The dual-radio architecture achieves the coverage of LoRa-only mesh (10+\,km) while maintaining BLE-class latency (32\,ms) and near-BLE energy efficiency for the majority of traffic. Neither single-radio architecture can match this combination: BLE-only lacks coverage; LoRa-only sacrifices latency, energy, and scalability. The crossover point --- where dual-radio outperforms both alternatives --- occurs when inter-cluster traffic is below $\sim$30\% ($\alpha < 0.30$), which is the common case in emergency scenarios with locality bias $\beta > 0.5$.

\begin{table}[!t]
\centering
\caption{Architecture Comparison Summary}
\label{tab:comparison}
\begin{tabular}{@{}lccc@{}}
\toprule
\textbf{Metric} & \textbf{BLE-only} & \textbf{LoRa-only} & \textbf{Dual} \\
\midrule
Latency (2-hop) & \textbf{32\,ms} & 5.0\,s & \textbf{32\,ms} \\
Energy/msg (avg) & \textbf{0.26\,mJ} & 92.5\,mJ & 2.5\,mJ \\
Max nodes & 15 & 50--80 & \textbf{250--562} \\
Coverage & 2.4\,km & \textbf{10+\,km} & \textbf{10+\,km} \\
Battery (500\,mAh) & \textbf{3.2\,d} & 1.6\,d & 2.3--3.2\,d \\
\bottomrule
\end{tabular}
\end{table}

\section{Related Work}

Meshtastic~\cite{ref_meshtastic} uses LoRa-only managed flooding with SNR-based contention; deployments reveal 80\% PDR and 50--80 node scalability limits~\cite{ref_wong_survey}. LoRaBLE~\cite{ref_loraBLE} integrates BLE and LoRa but with single-hop LoRa, static cluster assignment, and no mesh backbone. The Bluetooth Mesh SIG standard~\cite{ref_bt_mesh} (v1.1, 2023) added Directed Forwarding but still lacks Coded PHY support. O-AODV~\cite{ref_oaodv} and M-O-AODV~\cite{ref_moaodv} implement AODV on BLE mesh in simulation but without long-range capability. LEACH~\cite{ref_heinzelman} and HEED~\cite{ref_heed} established clustering theory for WSN but target single-radio sensor networks. Del-Valle-Soto~\textit{et~al.}~\cite{ref_delvalle} recently evaluated LoRa-BLE hybrid networks for jamming resilience but without mesh routing. Bluemergency~\cite{ref_bluemergency} provides BLE emergency mesh limited to $\sim$100\,m. Our architecture uniquely combines BLE advertising AODV with multi-hop LoRa backbone, formal traffic analysis, and implementation on COTS hardware.

\section{Discussion}

\subsection{Practical Deployment Considerations}

The architecture is designed for zero-configuration deployment: nodes power on, discover neighbors via BLE beacons, form clusters, and elect CHs within $\sim$6\,s. No provisioning server, key distribution, or manual configuration is required. This is critical for emergency scenarios where trained operators may not be available. The six-node prototype uses off-the-shelf Heltec Mesh Node T114 boards (retail $\sim$\$18/node as of 2025), making large-scale deployment economically feasible compared to military mesh radios costing \$10,000+ per unit.

\subsection{Limitations}

The analysis relies on the locality assumption ($\beta \geq 0.76$). In scenarios with high inter-cluster traffic ($\alpha > 0.30$), the LoRa backbone becomes a bottleneck. The current CH election (highest ID) includes a basic battery-threshold demotion mechanism but does not proactively optimize for energy balance; a residual-energy-weighted election would improve network lifetime fairness. Security is not addressed in the current protocol --- BLE advertising is unencrypted and vulnerable to spoofing. Adding AES-128-CCM (supported by nRF52840's CryptoCell-310 hardware accelerator in $\sim$15\,$\mu$s) is planned for the next firmware revision. Finally, the LoRa radio (SX1262) is present on the hardware but not yet activated in firmware; the analytical results are derived from datasheet parameters and validated LoRa performance literature rather than field measurements.

\section{Conclusion}

We presented a dual-radio BLE-LoRa hierarchical mesh with a formal traffic offloading model showing that emergency communication locality ($\beta \geq 0.76$) keeps 82--90\% of traffic on BLE, reducing energy by 79\% versus LoRa-only mesh. The architecture supports 250--562 nodes (10$\times$ Meshtastic's practical limit) with 10\,km+ diameter in 3.0\,KB RAM on commodity nRF52840$+$SX1262 hardware at \$18/node. Future work includes LoRa firmware activation and field evaluation, energy-aware CH rotation, lightweight encryption, and smartphone BLE integration for civilian usability.


\newcommand{\authorphoto}{\includegraphics[width=1in,height=1.25in,clip,keepaspectratio]{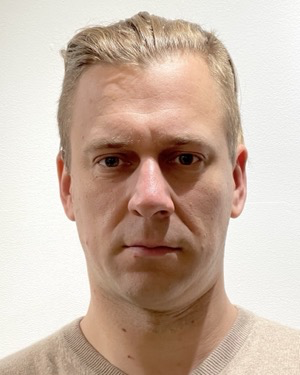}}
\begin{IEEEbiography}[\authorphoto]{Andrii Vakhnovskyi}
received the B.S. degree in computer engineering and the M.S. degree in systems engineering from the National Technical University ``Kharkiv Polytechnic Institute,'' Ukraine, in 2009 and 2011, respectively. He is the Founder and CEO of Biiper LLC, New York, NY, developing off-grid mesh communication systems, and IOGRU LLC, developing AI-driven IoT platforms for controlled environment agriculture. He is a Senior Member of ISA and a Member of IEEE. His research interests include mesh networking, embedded systems, BLE/LoRa protocols, and industrial IoT.
\end{IEEEbiography}


\begin{thebibliography}{99}

\bibitem{ref_fcc_maria}
FCC, ``Communications status report for areas impacted by Hurricane Maria,'' Dec. 2017; GAO-21-297, 2021.

\bibitem{ref_meshtastic}
N.K.~Suryadevara and A.~Dutta, ``Meshtastic infrastructure-less networks for reliable data transmission,'' in \textit{Proc. WiSATS}, LNCS, vol.~410, Springer, 2022.

\bibitem{ref_wong_survey}
Y.~Wong \textit{et~al.}, ``LoRa mesh networking: A survey,'' \textit{ACM Computing Surveys}, vol.~56, no.~8, 2024.

\bibitem{ref_bridgefy}
M.R.~Albrecht, R.~Eikenberg, and K.G.~Paterson, ``Breaking Bridgefy, again: Adopting libsignal is not enough,'' in \textit{Proc. USENIX Security}, 2022.

\bibitem{ref_fema_usar}
FEMA, ``US\&R field operations guide,'' 2022; NIST SP~1262, ``Emergency communications during Hurricane Maria,'' 2020.

\bibitem{ref_nordic_range}
Nordic Semiconductor, ``Testing long range (Coded PHY) with Nordic solution,'' 2019.

\bibitem{ref_sx1262}
Semtech, ``SX1261/62 long range, low power, sub-GHz RF transceiver,'' Datasheet Rev.~1.2, 2020.

\bibitem{ref_loraBLE}
L.~Leonardi \textit{et~al.}, ``LoRa-BLE multi-hop networks for IoT monitoring,'' \textit{Computer Communications}, vol.~190, pp.~204--218, 2022.

\bibitem{ref_ctc_survey}
T.~He \textit{et~al.}, ``Cross-technology communication for the Internet of Things: A survey,'' \textit{ACM Computing Surveys}, vol.~55, no.~4, 2022. DOI: 10.1145/3530049.

\bibitem{ref_aodv_rfc}
C.~Perkins, E.~Belding-Royer, and S.~Das, ``Ad hoc on-demand distance vector (AODV) routing,'' RFC~3561, Jul. 2003.

\bibitem{ref_bt_mesh}
Bluetooth SIG, ``Mesh profile specification v1.1,'' Sep. 2023.

\bibitem{ref_oaodv}
M.R.~Ghori \textit{et~al.}, ``Bluetooth mesh network performance analysis for IoT,'' \textit{Electronics}, vol.~10, no.~21, 2021.

\bibitem{ref_moaodv}
M.R.~Ghori \textit{et~al.}, ``Multipath optimized AODV routing for Bluetooth mesh,'' \textit{Sensors}, vol.~24, no.~3, 2024.

\bibitem{ref_heinzelman}
W.~Heinzelman, A.~Chandrakasan, and H.~Balakrishnan, ``Energy-efficient communication protocol for wireless microsensor networks,'' in \textit{Proc. HICSS}, 2000.

\bibitem{ref_heed}
O.~Younis and S.~Fahmy, ``HEED: A hybrid, energy-efficient, distributed clustering approach,'' \textit{IEEE Trans. Mobile Computing}, vol.~3, no.~4, pp.~366--379, 2004.

\bibitem{ref_delvalle}
C.~Del-Valle-Soto \textit{et~al.}, ``Adaptive jamming mitigation for clustered LoRa-BLE hybrid WSN,'' \textit{Sensors}, vol.~25, no.~6, 2025.

\bibitem{ref_bluemergency}
A.~Pozza \textit{et~al.}, ``Bluemergency: Distributed emergency messaging via BLE mesh,'' in \textit{Proc. IEEE DCOSS}, 2020.

\end{thebibliography}
\end{document}